\documentclass{aa}
\usepackage{psfig} 
\begin{document}
\thesaurus{11.06.2; 11.07.1; 11.09.2; 11.11.1; 11.19.2}
\title{Central Matter Distributions in Rich Clusters of Galaxies from z$\sim$0 
to z$\sim$0.5 \thanks{Based on observations collected at the European Southern
Observatory 
(La Silla, Chile) and at the CFH observatory (Hawaii), and on the POSS and the 
Cosmos Survey.} }

\author{C.~Adami \inst{1}, A.~Mazure \inst{1}, M.P.~Ulmer \inst{2}, 
C.~Savine \inst{1}}
\institute{LAM, Traverse du Siphon, F-13012 Marseille, France
\and NU, Dearborn Observatory, 2131 Sheridan, 60208-2900 Evanston, USA}
\offprints{C.~Adami} 
\date{Received date; accepted date} 
\maketitle 
\markboth{Central Matter Distributions in Rich Clusters of Galaxies from 
z$\sim$0 to z$\sim$0.5}{} 

\begin{abstract} 
We have analyzed the galaxy number density and luminosity density profiles
of rich clusters of galaxies from redshifts  z$\sim$0 to z$\sim$0.5.  
We show that the luminosity profile computed with bright galaxies 
(M$_R < -21$) is significantly cusped in the center of the clusters, whatever
the redshift. This is in agreement with the dark matter profiles predicted by 
numerical simulations. The galaxy number density profile for the bright 
galaxies is fitted equally well with a core model or a cusped model. In 
contrast, the luminosity and the galaxy number density profiles of the fainter 
galaxies are significantly better fitted by a core rather than a cusp model.  
We did not detect any statistically significant different fits when applied to 
data in range from z$\sim$0 to z$\sim$0.5. The difference in profile between 
faint and bright galaxies may be due to the rapid 
(relative to the age of the universe at z=0 versus z=0.5) destruction of 
the faint galaxies by tidal forces and merging events in the denser central 
regions of the clusters. This process could erase the cusp by turning faint 
galaxies into diffuse light. In this case, the galaxies (with a cusp visible 
in the bright galaxy number density and mainly in luminosity profiles) would 
trace the total mass distribution. 

\end{abstract}

\begin{keywords} 
{ 
Galaxies: clusters: general; Cosmology: large-scale structure of Universe
} 
\end{keywords}

\section{Introduction}

Clusters of galaxies are key cosmological probes. They are one of the main
products of the hierarchical models and the largest virialized structures in
the Universe. Recent studies (e.g. Navarro et al. 1997) argue that whatever 
the cosmological model, a universal profile 
(NFW profile) with a cusp at the center should describe the cluster dark 
matter profile, in contrast to the beta-model (King 1962) which has been used 
from many years (cf Sarazin 1986). The shape of the galaxy number
and luminosity density profiles 
(and their dependence on redshift) relates directly to the physical processes 
acting at the center of the clusters and we propose to investigate in this way 
these processes. While the dark matter profile (close to the total mass
profile) seems to be very concentrated (Navarro et al. 1997), the galaxy 
distribution is nearly flat in the center of the nearby clusters (Adami et al. 
1998: ENACSVII).  Furthermore, the X-ray gas distribution cannot be used to 
distinguish between a flat and a cusped model except for the nearby clusters
(e.g. Durret et al. 1994) due to a lack of resolution with Rosat, Einstein and 
Asca and due to cooling-flows.

In this paper we contribute to the effort to solve 
this puzzle by studying the galaxy number 
density profile and the galaxy luminosity 
density profile (the sum of the luminosity inside the galaxies; we will
consider the diffuse light only at the end of the paper). This approach is 
clearly not new (e.g. Mazure et al. 1986), but we used here very large, 
homegeneously analyzed, samples. The use of all the optical wave band galaxy 
light allows us to take into account merging events. This method 
enables us to recover "erased" cusps as it is the light of merged galaxies 
which is counted, not the number of galaxies. 
The best way to investigate the mass profile would be to compute this
profile via the velocity dispersion profile (e.g. Carlberg et al. 1997a
and b, Biviano et al. 2001), the X-ray temperature profile, or gravitational 
lensing, but these techniques require large amounts of data, and therefore, it 
is very difficult to obtain large samples of clusters. Moreover, galaxy 
luminosity density profiles for early type galaxies are claimed (e.g. Kaiser 
1999) to be similar to the mass profile in clusters. This is an important
question as galaxies would be sufficient to trace the mass profile. We will 
use in this paper homogeneous samples of thousands of galaxies and tens of 
clusters to compare the projected galaxy number density and luminosity profiles, and
the simulated dark matter profiles (e.g. Navarro et al. 1997).

In Section 2, we describe the samples. In Sections 3 and 4, we describe the 
methods we used and the profiles we generated. Section 5
discusses the results and Section 6 gives a summary of the results.

To relate values of z to distance and to be in agreement with Adami et al. 
(1998, ENACSVII), we have taken H$_0$ = 100 km.s$^{-1}$Mpc$^{-1}$, q$_0$ = 0 
and $\Lambda = 0$.

\section{Samples}

\subsection{Low redshift sample}

We used the COSMOS composite cluster which is decribed in detail in ENACSVII,
but we briefly enumerate the main points here: we compiled photometric COSMOS
data (Heydon-Dumbleton et al. 1989) for 77 of the 107 ENACS clusters 
(z$\leq$0.1). We have adopted a limiting magnitude of $b_j$ = 20. 

We have selected the clusters with:

\begin{enumerate}
\item a low level of substructure 
(both visually and with a Dressler-Shectman test) in a 10 core radii on a side 
square ($\sim$1200 kpc h$^{-1}$), with more than 10 known redshifts in the 
main group (to avoid superposition effects); 

\item a redshift lower than 0.1 

\item and a
converging solution of the fit of analytical density profiles. 

\end{enumerate}

These selection criteria lead to a sample of 29 homogeneously selected clusters.
With this sample we built a composite cluster with about 5000 galaxies. For 
each of the 29 clusters, we determined the center by a Maximum Likelihood fit 
and we confirmed the results with other estimators (X-Ray centers, cD galaxies,
see also Ulmer et al. 1992). For each galaxy, the distance from the center has 
been scaled both with the core radius of the cluster and the r$_{200}$ (radius 
where the density is equal to 200 times the critical density) in order to take
into account that different clusters have different sizes (see ENACS VII for
a discussion of this point). We
have taken the elongation of the clusters into account by 'circularizing' the
individual galaxy distributions by increasing all projected distances
orthogonal to the major axis, thus 'expanding' the distribution parallel to
the minor axis by a factor deduced from the ellipticities of individual 
clusters. This correction is important because the superposition of galaxy
distributions with randomly distributed orientations will cause outer
densities to be underestimated with respect to the inner ones, which is
producing an artificial cusp (cf ENACS VII).

\subsection{High redshift sample}

We used a sample of 7 clusters from the COP (Adami et al. 2000a, Holden
et al. 2000) and CNOC (e.g. Carlberg et al. 1996) surveys selected with
the same conditions as the low redshift sample:
\begin{enumerate}

\item low level of substructures in a 10 core radii square ($\sim$1200
kpc) from visual analysis

\item more than 10 redshifts in the main group

\item converging solutions of the fit of analytical galaxy number 
and luminosity
density profiles (this excludes 2 clusters: MS1008 and MS1224).
\end{enumerate}
This sample is described in Table 1: the redshift range of this sample of 
distant clusters is [0.33;0.55] (mean redshift = 0.42). We have built a 
composite cluster of 459 galaxies with these 7 clusters, correcting for
ellipticity and orientation of each real cluster, in a similar manner as
we did for the nearby sample.

We note that MS1358 is regarded in the CNOC papers as a complex cluster.
However, we used in this paper only the central regions of this cluster in
order to focus on the central cluster shape: we limited this cluster to
a region of $\sim$160'' radius. With this restriction, MS1358 can be included in 
our sample. We stress, however, that using larger areas will induce 
significant levels of substructures.

\section{Methods and Results}
\subsection{Low redshift galaxy number density and luminosity profiles}

The individual parameters (with individual clusters) of the low redshift 
sample profiles are described 
in Adami et al. (1998: ENACSIV). They have been computed with a 
Maximum Likelihood fit.  We can easily generalized the same code to fit a 
luminosity profile for the composite clusters. We have weighted each galaxy
with its luminosity (assuming all the galaxies at the cluster redshift): 
$Lum_k$. 
Similarly to ENACSVII, the probability that the assumed profile `produces' a 
galaxy in position (x$_k$,y$_k$) is $\sigma (x_k,y_k)$ (with k=1...N). The
combined probability that the assumed profile will produce galaxies in the 
positions (x$_k$,y$_k$) that they actually have is:

L=$\prod\limits_{k=1}^N Lum_k  \times \sigma (x_k,y_k)$

The best fit model is found from a maximization of L. We have chosen, instead,
to compute -ln(L) with:

ln L = $\sum\limits_{k=1}^N ln (Lum_k  \times \sigma (x_k,y_k))$

The best model (which produces the best fit) is found from a minimization of 
-ln(L). We minimize this value by using the MINUIT package (e.g. ENACSVII for 
details).

To be able to quantify the degree of cuspiness at the center of the clusters, 
we fitted in ENACSVII both a projected beta-model:

$\mu (r)=\mu _0(1/(1+ (\frac{r}{r_c})^2)^\beta +\mu_b$

and a cusped profile:

$\mu (r)=\mu_0( 1/(\frac{r}{r_c}(1+\frac{r}{r_c}) ^2)) ^{\beta }+\mu_b$

For the luminosity profiles, $\mu$ is the surface brightness, which replaces
the surface density $\sigma$ of the formulae in ENACSVII.
These models, fitted to the individual clusters, have 7 free parameters: two 
parameters for the position of
the center (x$_0$,y$_0$), two parameters to describe deviations from
symmetry (ellipticity e and position angle $\phi$, included in r), two 
parameters that specify the profile (r$_c$ and $\beta$) and the
background density $\mu_b$ (assumed constant within the aperture of
each cluster).

To be able to compare the fit quality of the two models (cusped profile and
beta-model), we use the logarithm 
of the ratio of the Maximum Likelihood values for each fit as explained in 
ENACSVII. This value has a $\chi ^2$ distribution (Meyer 
1975) which we then use to determine the statisticial significance level
of the best fit of one profile type versus the other.

Moreover, we have checked the consistency between the center of the
clusters determined by using the galaxies (see ENACSVII) and the center 
determined by using the galaxies weightened with their luminosity. We find a 
shift of only (32 $\pm$ 21) kpc. This shift is not large 
enough to erase a cusp (e.g. ENACSVII Fig. 8) and we have assumed the 
centers using only the galaxy number density.

\subsection{High redshift galaxy number density and luminosity profiles}
             
We used a similar approach as for the low redshift sample, but with an 
additional step to take into account the high background level.  This is
because it is nearly impossible to measure redshifts for all galaxies of all 
the clusters in our high redshift sample due to the large amount of observing 
time
needed.  Therefore we used the red sequences in the Color Magnitude Relations 
(CMR hereafter) to reject many of the field galaxies along the lines of sight
(e.g. Yee et al. 1999).

\begin{table*} 
\caption[]{High redshift sample with cluster name, selection of the
cluster galaxies (color interval or redshift), redshift, characteristic radius 
and $\beta$ for the beta-model (galaxy number density profile: dens. gal.), 
characteristic radius and $\beta$ for the cusped model (galaxy number 
density profile: dens. gal.), characteristic radius and $\beta$ for the 
beta-model (galaxy luminosity profile: dens. lum.), characteristic radius 
and $\beta$ for the cusped model (galaxy luminosity profile: dens. lum.) and 
best fit (galaxy number and luminosity density profiles). The quoted errors are 
at the 1-$\sigma$ level. No error means that we were not able to get a 
reliable estimate for this error.}
\begin{flushleft} 
\begin{tabular}{ccccc} 
\hline 
\noalign{\smallskip} 
Name & Selection & z \\ 
$r_c$ dens. gal. beta-model & $\beta$ dens. gal. beta-model & $r_c$ dens. gal. 
cusped & $\beta$ dens. gal. cusped & best fit dens. gal. \\ 
$r_c$ dens. lum. beta-model & $\beta$ dens. lum. beta-model & $r_c$ dens. lum. 
cusped & $\beta$ dens. lum. cusped & best fit dens. lum. \\ 
\hline 
\hline 
\noalign{\smallskip} 
MS0302 & g-R: [1.00,1.80] & 0.42 \\
63$\pm$12 kpc & 0.93$\pm$0.06 & 93$\pm$32 kpc & 0.52$\pm$0.04 & beta-model: 
not significant \\
62$\pm$12 kpc & 0.94$\pm$0.09 & 94$\pm$29 kpc & 0.52$\pm$0.05 & beta-model: 
not significant \\
\hline 
MS0451 & g-R: [1.25,1.85] & 0.54 \\
121$\pm$32 kpc & 0.87$\pm$0.17 & 209$\pm$108 kpc & 0.53$\pm$0.07 & beta-model: 
not significant \\
117$\pm$29 kpc & 0.88$\pm$0.16 & 210$\pm$120 kpc & 0.53$\pm$0.08 & beta-model: 
not significant \\
\hline 
MS1358 & redshift & 0.33\\
117$\pm$26 kpc & 0.98$\pm$0.09 & 455$\pm$97 kpc & 0.81$\pm$0.05 & cusped: not 
significant \\
119$\pm$29 kpc & 0.99$\pm$0.08 & 469$\pm$95 kpc & 0.81$\pm$0.05 & cusped: not 
significant \\
MS1358 & g-R: [1.02,1.50] & 0.33 \\
95$\pm$22 kpc & 0.91$\pm$0.07 & 516$\pm$172 kpc & 0.83$\pm$0.06 & cusped: not 
significant \\
101$\pm$13 kpc & 0.92$\pm$0.04 & 509$\pm$151 kpc & 0.83$\pm$0.05 & cusped: not 
significant \\
\hline 
MS1621 & g-R: [0.85,1.85] & 0.43 \\
67 kpc & 0.80 & 735 kpc & 0.86 & beta-model: not significant \\
68 kpc & 0.81 & 737 kpc & 0.85 & beta-model: not significant \\
\hline 
PDCS16 & V-I: [1.35,2.11] & 0.40 \\
299$\pm$39 kpc & 0.99$\pm$0.20 & 882 kpc & 0.68 & beta-model: not significant \\
303$\pm$36 kpc & 0.98$\pm$0.18 & 885 kpc & 0.67 & beta-model: not significant \\
\hline 
PDCS30 & V-I: [1.10,1.90] & 0.33 \\
189 kpc & 1.04 & 302$\pm$67 kpc & 0.53$\pm$0.04 & beta-model: 85$\%$ level \\
189 kpc & 1.01 & 227$\pm$69 kpc & 0.46$\pm$0.03 & beta-model: 85$\%$ level \\
\hline 
PDCS62 & V-I: [1.24,2.20] & 0.46 \\
196$\pm$10 kpc & 1.01$\pm$0.26 & 284$\pm$73 kpc & 0.53$\pm$0.04 & beta-model: 
85$\%$ level \\
192 kpc & 1.02 & 282$\pm$71 kpc & 0.53$\pm$0.03 & beta-model: 85$\%$ level \\
\noalign{\smallskip} 
\hline	    
\normalsize 
\end{tabular} 
\end{flushleft} 
\label{} 
\end{table*}

The color intervals we used are given in Table 1. We chose these intervals
manually in order to match the prominent structures in the CMR
of each line of sight.  We have confidence that we made good choice of
intervals as these intervals
are very similar to the 
theoretical predictions of Kodama $\&$ Arimoto (1997). 

We give in Table 1 the best fit values of $r_c$ and $\beta$ for the 7 high
redshift clusters, the 2 analytical models (beta-model and cusped profile) and 
the galaxy number and luminosity density profiles parameters ($r_c$ is ranging from 
63 to 299 kpc and $\beta$ from 0.8 to 1.04 for a beta-model). We give also the 
fit quality: which analytical model fits best, core or cusp?

We now examine whether that these values could have been affected by the
following bias: the CMR rejects not only field galaxies but also the late type 
galaxies that are cluster members. Since these galaxies are preferentially 
located in the outskirts of clusters (e.g. Adami et al. 1998a), this could
affect the density profiles.
 
-- At low redshift, this is not a serious concern as more than 
75$\%$ of the cluster galaxies in a 1200 $h^{-1}$kpc square are early type
objects (Adami et al. 1998a).

\begin{figure} 
\vbox 
{\psfig{file=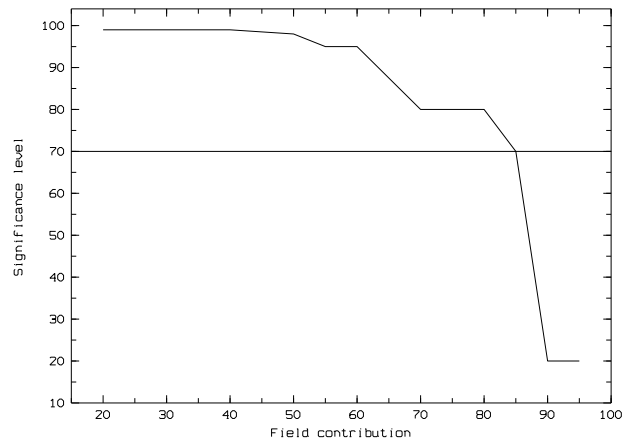,width=9.0cm,angle=270}} 
\caption[]{Significance level of the core/cusp discrimination in our
simulations (y-axis: percentage) as a function of the field contribution to 
the total number of galaxies along the line of sight (x-axis: percentage). The 
horizontal line is the 70$\%$ significance level.} \label{} 
\end{figure}

-- For the higher redshift clusters, however, the spiral fraction
increases (e.g. Dressler et al. 1997) and can contribute for up to 60$\%$ in
the outer parts of the clusters (Ellingson et al. 2000). As we want to focus 
on the central cluster shape, we limited ourselves to a region of 
$\sim$0.5$r_{200}$ (5 core radii). In this region, the field-like galaxy
contribution is lower: $\sim$30$\%$. We have tested, however, the 
significance of the effect by removing these galaxies when we fit the density 
profiles. We used the CNOC cluster MS1358 because redshifts are available for 
nearly all the galaxies along the line of sight. We can, therefore, select the 
cluster member galaxies on a firm basis.  We see in Table 1 that using the CMR 
or the redshift selection does not affect in a significant manner
the fitted parameters: these parameters are similar at the 1-$\sigma$
level: for a beta-model $r_c$ is ranging from 117 to 95 kpc and $\beta$ from
0.91 to 0.98. This is not a definitive test as we are using a single cluster,
but the results are still very suggestive.

Using the CMR seems, therefore, a good way to remove field galaxies in order
to discriminate between cusped profiles and profiles with a core. However,
as we go to faint magnitudes, the uniform field contribution becomes
stronger and stronger. This field contribution is between 50$\%$ and 75$\%$
for the clusters in our sample (estimated from the redshift catalogs of these
cluster lines of sight). This could have the effect that the cluster density
profiles will $appear$ flatter for faint galaxies.  

--This does not affect our results for nearby clusters (we do not sample the 
clusters at magnitudes fainter than R=19, where the CMR becomes to be
less contrasted compared to the field contribution: Adami et al. 2000b). 

--This does not affect our results for distant clusters (z$\sim$0.4) 
for the brightest bin of Table 3 (magnitudes brighter than R=19).

--This could have an influence for the 3 faintest bins of 
the distant composite cluster. However, the goal of this paper is to
discriminate between density profiles with a cusp or with a core. In order to
quantify the probability to have a cusped profile mis-interpreted as a profile
with a core due to a high field contribution, we performed simulations. We
generated a circular artificial cluster profile with a cusp in the center (as 
described in Section 3.1 with $r_c$=200kpc and $\beta$=0.67) and we added to 
this cluster a uniform field contribution with various galaxy densities. We 
computed finally the relative fit quality between a profile with a core and 
with a cusp. For field contributions as high as 80$\%$ of the total number of
galaxies along the line of sight, a model with a cusp is prefered at the
80$\%$ significance level. For contributions equal to 85$\%$, a model with
a cusp is prefered at the 70$\%$ significance level.
For higher field contributions, we are not able to
discriminate between a profile with a core or with a cusp (a model with
a cusp is still prefered but with significance levels lower than 65$\%$, which
is not significant). The results are plotted in Fig. 1. This means that even
with a high field contribution, we are still able to discriminate between 
profiles with a core or with a cusp
because the field contribution along the lines of sight we used is not higher
than 80$\%$. Our results are, therefore, valid regarding the
core/cusp discrimination.

\subsection{Analysis of the results}
                          
\begin{table*} 
\caption[]{Galaxy number density and luminosity profile parameters 
(characteristic 
radius $r_c$ and  slope $\beta$) for nearby and distant clusters. The values 
for 
the individual cluster galaxy number density profiles are the mean and the 
1-$\sigma$ 
uncertainty between the individual 
fits. The values for the composite clusters are the fitted values. No error 
means that we were not able to get a reliable estimate for this error.}
\begin{flushleft} 
\begin{tabular}{cccccc} 
\hline 
\noalign{\smallskip} 
DENSITY & $r_c$ beta-model & $r_c$ cusped & $\beta$ beta-model & $\beta$ 
cusped & best fit \\ 
\hline 
\hline 
\noalign{\smallskip} 
Nearby sample (z$\sim$0.07) & $<$128$>$$\pm$88 kpc & $<$292$>$$\pm$191 kpc & 
$<$1.02$>$$\pm$0.08 & $<$0.61$>$$\pm$0.05 & beta-model \\ 
(Individual clusters) & & & & & Not significant \\
\hline 
Distant sample (z$\sim$0.4) & $<$147$>$$\pm$79 kpc & $<$431$>$$\pm$268 kpc & 
$<$0.94$>$$\pm$0.08 & $<$0.64$>$$\pm$0.14 & beta-model \\ 
(Individual clusters) & & & & & Not significant\\
\hline 
Nearby sample (z$\sim$0.07) & 89$\pm$5 kpc & 318$\pm$34 kpc & 1.00$\pm$0.02 & 
0.56$\pm$0.01 & beta-model \\ 
(Composite cluster) & & & & & 99$\%$ level \\
\hline 
Distant sample (z$\sim$0.4)  & 119$\pm$8 kpc & 215$\pm$9 kpc & 1.08$\pm$0.05 & 
0.65$\pm$0.04 & beta-model \\ 
(Composite cluster) & & & & & 85$\%$ level \\
\hline 
\hline 
\hline 
LUMINOSITY & $r_c$ beta-model & $r_c$ cusped & $\beta$ beta-model & $\beta$ 
cusped & best fit \\ 
\hline 
\hline 
\noalign{\smallskip} 
Nearby sample (z$\sim$0.07) & 101$\pm$7 kpc & 281$\pm$41 kpc & 1.01$\pm$0.03 & 
0.58$\pm$0.03 & beta-model: not significant \\ 
\hline 
Distant sample (z$\sim$0.4)  & 119$\pm$9 kpc & 232 kpc & 1.08$\pm$0.04 & 
0.66 & beta-model: 85$\%$ level \\ 
\noalign{\smallskip} 
\hline	    
\normalsize 
\end{tabular} 
\end{flushleft} 
\label{} 
\end{table*}

--We compare the characteristic parameters $r_c$ and $\beta$ of the distant
clusters with those of the nearby sample (ENACSVII). The results are 
summarized in Table 2 and we give the description of these results: we 
used first the individual clusters. We do not see a statistically significant 
evolution with redshift: the agreement between z$\sim$0 and z$\sim$0.4 
fits is consistent at the $2\sigma$ level (for a beta-model, individual 
clusters and galaxy number density, $r_c \sim 130kpc$ and $\beta \sim 1$). 
If we compare 
now the fitted parameters $r_c$ and $\beta$ of the galaxy number and 
luminosity density profiles (regardless of the redshift), they also agree at the 
2-$\sigma$ level (for a beta-model, composite clusters and luminosity 
profiles, $r_c \sim 110kpc$ and $\beta \sim 1.05$).

--As for the nearby sample (e.g. ENACSVII), the cluster galaxy distribution of 
the  distant cluster sample is better fitted with beta-models than with 
cusped profiles. Although the fit is better, it is not significantly
better than a 75$\%$ significance level.

--To improve the statistical significance of the fits in order to distinguish
between  cusped profile and beta-profile fits, we used also composite 
clusters. The method used to build these composite clusters is described
in Section 2.1. We found
that the behavior of the fits was similar for the distant and nearby
samples: the bright galaxies better followed a cusped luminosity 
profile, while the faint galaxy luminosity and galaxy number density 
profiles were 
better fitted with beta-models. This is also illustrated for the nearby
sample with Fig. 2. We see that the bright galaxy luminosity density profile 
seem to be more peaked than for the fainter galaxies.

\begin{figure} 
\vbox 
{\psfig{file=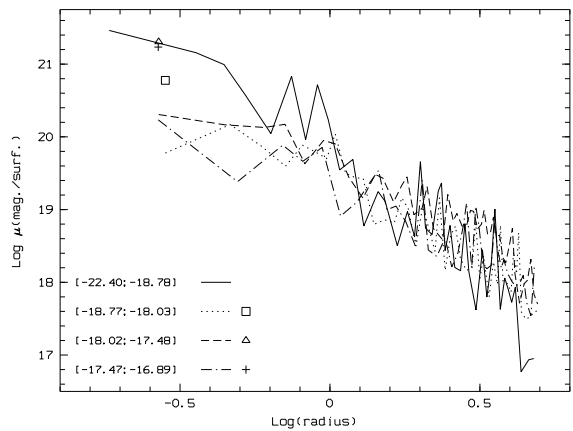,width=9.0cm,angle=270}} 
\caption[]{Illustration of the luminosity profiles for the r$_c$ scaled
nearby
composite 
clusters (assuming $r_c$$\simeq$100 kpc) and the four luminosity ranges. The 
r$_c$ axis is in logarithmic units of r$_c$. Each bin has the same number
of 
galaxies and the y values are rescaled to the same integral.
The 3 symbols represent the value of the central bin with the
diffuse light correction applied to the luminosity density profiles listed
in the figure next to the relevant symbol.} \label{} 
\end{figure}

\begin{figure} 
\vbox 
{\psfig{file=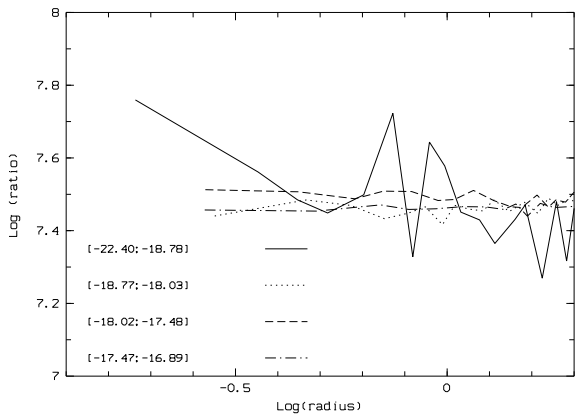,width=9.0cm,angle=270}} 
\caption[]{Log. of the ratio of the luminosity profiles and the galaxy number 
density 
profiles for the r$_c$ scaled nearby composite clusters (assuming 
$r_c$$\simeq$100 kpc) and the four magnitude ranges.} 
\label{} 
\end{figure}

\section{Cores or Cusps: Comparison with Literature}

We have analyzed in this paper a very large sample of clusters and we have 
shown that we have a better fit for the galaxy number 
density if we used a model with 
a core rather than with a cusp (except for the bright galaxies). 
This is apparently contrary to the study of Carlberg et al. (1997a) 
which favored 
a model with a cusp for the galaxies of the  CNOC clusters. We have shown in
ENACSVII, however, that the results of Carlberg et al. are probably explained 
because Carlberg et al. did not correct for ellipticity of the cluster 
profiles, and more importantly, they only considered bright galaxies. 

We have shown that if we limit ourselves to bright galaxies only (see Tab. 3), 
there is no statistically  significant difference between a model with a cusp 
or with a core for the galaxy number density. This result is in good agreement, for 
example, with the work of Biviano et al. (1996) on the Coma cluster.

For another comparison we remark that Durret et al. (1994) showed that,
using the X-Ray surface brightness profile 
of 12 clusters (based on ROSAT data), a model with a core does not fit 
significantly better the observations than a model with a cusp. 
                                                           
If we fit the luminosity density profiles for all galaxies with a beta-model 
and a cusped model, the
core model is still preferred, but the difference is not significant
over the redshift range from 0 to 0.5.
For a comparison, we have 
plotted in Fig. 3 the ratio of the galaxy luminosity density to the galaxy 
number density (for the 4 luminosity bins). 
We see that the ratio is constant for the faintest bins, while
we have an increasing ratio for the brightest bin close to the 
center, inside $\sim$50 kpc. As we have noted in previous sections, there
seems to be a difference between the distribution of bright and faint galaxies
and also between galaxy number density profile and the luminosity density
profile.
\begin{table*} 
\caption[]{Best fit for the different models and composite clusters.}
\begin{flushleft} 
\begin{tabular}{ccc} 
\hline 
\noalign{\smallskip} 
Nearby composite cluster & Luminosity density profile & Galaxy number 
density profile \\ 
\hline 
\noalign{\smallskip} 
All galaxies, $r_c$ scaled & beta-model better than cusp: not significant
& beta-model better than cusp 99$\%$ level\\
All galaxies, $r_{200}$ scaled & beta-model better than cusp: not significant
& beta-model better than cusp 99$\%$ level\\
$b_j$ [-22.4;-18.78], $r_c$ scaled & cusp better than beta-model 95$\%$ level
& beta-model better than cusp: not significant \\
$b_j$ [-18.77;-18.03], $r_c$ scaled & beta-model better than cusp 95$\%$ level
& beta-model better than cusp 95$\%$ level\\
$b_j$ [-18.02;-17.48], $r_c$ scaled & beta-model better than cusp 75$\%$ level
& beta-model better than cusp 95$\%$ level\\
$b_j$ [-17.47;-16.89], $r_c$ scaled & beta-model better than cusp 75$\%$ level
& beta-model better than cusp 75$\%$ level\\
\noalign{\smallskip} 
\hline 
\noalign{\smallskip} 
Distant composite cluster & Luminosity density profile & Galaxy number 
density profile \\ 
\hline 
\noalign{\smallskip} 
All galaxies, $r_c$ scaled & beta-model better than cusp 85$\%$ level
& beta-model better than cusp 85$\%$ level \\
R [-22.;-21.], $r_c$ scaled & cusp better than beta-model 95$\%$ level
& beta-model better than cusp: not significant \\
R [-21.;-20.5], $r_c$ scaled & beta-model better than cusp 95$\%$ level
& beta-model better than cusp 95$\%$ level\\
R [-20.5;-20.], $r_c$ scaled & cusp better than beta-model: not significant
& cusp better than beta-model: not significant \\
R [-20.;-19.], $r_c$ scaled & beta-model better than cusp 70$\%$ level
& beta-model better than cusp 70$\%$ level\\
\noalign{\smallskip} 
\hline	    
\normalsize 
\end{tabular} 
\end{flushleft} 
\label{} 
\end{table*}
It seems that two different shapes apply.  The faint galaxies exhibit a
core. They have a distribution in agreement with a beta-model. The bright 
galaxies (at least for the luminosity profile) show a cusp. This would be in
agreement with the dark matter profiles of the simulations of Navarro 
et al. (1997). As suggested by Kaiser (1999), the bright galaxies would, 
therefore, trace the cluster mass.

\section{Discussion}

We have presented different arguments favoring either a core or cusp in the
last sections. The galaxy number density profiles do not exhibit a cusp, and the 
bright galaxies cannot be used to distinguish between core and cusp models.
On the contrary, the luminosity profiles can be used to produce a cusp or a
core depending on whether just bright or faint galaxies are used to generate
the luminosity density profile. This leads to suggest 
different scenarios for the evolution of the bright and faint objects:

- the bright galaxies luminosity profile could be cusped at the formation 
epoch of the clusters. It could also become cusped via evolutionary processes, 
such as a segregation process (bright galaxies in the center of the clusters).

- the faint galaxies, which exhibit a core, 
could originally have been in a cusped distribution, but this distribution 
erased by environmental effects like tidal disruption or merging events. 
Galaxies near the cluster center and on radial orbits with low angular momentum 
would have to pass right through the giant ellipticals and would be eventually 
swallowed up or disrupted in the process. This can lead to a lower 
density of faint galaxies near the cluster center.  

\subsection{Segregation effects.}

We examine if the luminosity profile cusp for bright galaxies is
due to a segregation effect.  In Adami et al. (1998a) it was shown 
that the elliptical galaxies are the brightest objects in a cluster. Moreover, 
they are more concentrated in the cluster centers than the other 
morphological types.  This explains probably why we are not able to 
distinguish between a core and a cusp when we examine the galaxy number
density 
profile of the bright galaxies (ENACSVII): bright galaxies are mainly
elliptical and elliptical galaxies are mainly in the cluster centers.

To confirm this effect, we have removed the galaxies brighter than
b$_j=-21$ from the brightest bin of the nearby sample (28 galaxies). This
removes the cD-like galaxies. When re-doing the fit of a beta-model
and a cusped profile for the bright galaxy bin with the cDs
removed, we still prefer a model 
with a cusp,
with, however, at a lower significance level of 90 $\%$ (instead
of 95 $\%$). This effect does not appear, however, to be dominant.

We now discuss the difference between the galaxy number density and the
luminosity profiles. A cusp in the galaxy number 
density profile could be erased
by merging events, leading to a core. However, the luminosity profile 
of the bright galaxies has a cusp, but the same profile for the faint galaxies 
has a core.  We discuss a scenario that would explain this result in the next
subsection.

\subsection{Destruction of the faint galaxies in the center of the clusters?}

Suppose that the
formation mechanism of the clusters does not initially produce a core, but
only a cusp for both bright and faint galaxies, as in the simulations 
(e.g. Navarro et al. 1997). This cusp could be "erased" in the galaxy number
density
profile by merging events and "revealed" in the luminosity profile for the 
bright galaxies. We must now explain why the luminosity profile for the 
faint
galaxies has no cusp. This could be explained if we consider the other sources
of galaxy destruction, beside the merging events: the tidal disruptions.
Tidal forces in the cluster center disrupt part of the faint galaxies, turning
them into diffuse light, while the bright ones (more massive and more robust) 
are conserved. This explanation was, for example,  proposed by Merrit (1984). 
By
using simulations of the dynamical evolution 
of the cluster core, he predicted a tidal radius (lowest value for the
size of a not tidally disrupted galaxy) of about 15 $h^{-1}$ kpc.  
The bright galaxies, 
like cD galaxies, are significantly larger. The cusp could be erased in this 
way only for the faint galaxies which are generally smaller. More recent
simulations by Moore et al. (1998) show the same trend
for the dwarf galaxies to be disrupted in the center of the clusters.
Such a destruction of the faint galaxies has been also proposed on 
observational bases for example by Secker et al. (1997) or
Gregg $\&$ West (1998). These last authors, using deep photometry of the Coma 
cluster, propose the disruption of the faint galaxies as an explanation of the
lack of dwarf galaxies in the core of this cluster (e.g. also Adami et al. 
2000b). 

A way to investigate this possibility is, for example, to search for 
cusped profile in the diffuse light. Gregg $\&$ West (1998) give 
an approximation of the total luminosity lost for the galaxies by disruption 
and turned into diffuse light in the center of the Coma cluster. This value 
is about 20 $\%$ of the luminosity of a cD galaxy. This contribution is around
50 $\%$ of the luminosity sum of all galaxies fainter than $-18.77$ in the 
central bin (where the cusp is significant for the bright galaxies) of our 
nearby composite cluster. After rescaling to the total luminosity
to get units coherent with Fig. 2, we see on this figure that such a 
contribution is large enough to enhance the three faint luminosity profiles, 
consistently with a cusped model. 

A more quantitative study has been done for this same cluster by Bernstein et 
al. (1995). Using very deep images of a cluster-central area of 
7'$\times$7', they conclude that the faint galaxy luminosity profile is
flat or decreasing in the central 40 kpc (see also Adami et al. 2000b for
a spectroscopic survey of this area), while the luminosity profile of the 
brighter galaxies is peaked. Moreover, considering now the diffuse
luminosity profile (not associated with visible galaxies), they also found a 
peaked profile. If we argue that the 
cusp of the faint galaxy luminosity 
profile is erased by tidal disruption of part of these galaxies,
we should recover at least partially this shape in the diffuse light 
profile. It is exactly what Bernstein et al. saw in the Coma cluster.
                 
\subsection{Destruction time scale?}
                                   
We demonstrated that the cluster profiles have the same central shape from 
z$\sim$0 to z$\sim$0.5. This
would imply that the time scale of the destruction processes described above 
(merging + disruption) would be relatively short. This is because all the 
faint galaxies accreted between z$\sim$0 and z$\sim$0.5 do not contribute to 
the reconstruction of  a cusp. The in-falling galaxies seem to be continuously 
destroyed, in a time close to the virialization time which is approximately 
1.5 Gyear (e.g. Sarazin, 1986) for faint galaxies. This estimate is, for 
example, in good agreement with the estimates of the time needed to disrupt a 
low surface brightness galaxy in the simulations of Calcaneo-Roldan et al. 
(2000).
An alternative explanation in agreement with the observations is that the 
accretion rate could be already very low at z=0.5 (e.g. Ellingson et al. 
2000).

\section{Summary}

We have shown in this article that, on one hand, the galaxy luminosity 
density profile is cusped for the bright galaxies from z$\sim$0 to 
z$\sim$0.5 and exhibits a core for the faint galaxies. On another hand, the
galaxy number density profile has a core (not significant for the bright galaxies).
This could be understood if we assume a cusped profile for all the galaxies in
agreement for example with the simulations of Navarro et al. (1997). The cusp
could be erased for the galaxy number density profiles by merging effects and for the
luminosity profile of the faint galaxies with the destruction of these 
galaxies by tidal forces. This seems to be confirmed by two 
observational studies of the Coma cluster (Gregg $\&$ West 1998 and 
Bernstein et al. 1995). 

If this model is confirmed, this would imply that clusters of galaxies mass
profiles are well traced by the bright galaxy luminosity density profiles.

\begin{acknowledgements}

{AC acknowledges the Dearborn Observatory staff for their hospitality 
during his postdoctoral fellowship and J. Mohr for useful discussions.}

\end{acknowledgements}

\vfill 

\end{document}